\documentclass[preprint]{aastex62}

\usepackage{subfigure}
\usepackage{url}
\usepackage{hyperref}
\usepackage{epsfig}
\usepackage{graphicx}
\usepackage{sidecap}
\usepackage{hanging}
\usepackage{color}
\usepackage{multirow}
\usepackage{enumerate}
\usepackage{natbib}
\usepackage{amssymb}
\usepackage{amsmath}

\newcommand{\sci}{Science}
\newcommand{\jatis}{JATIS}

\newcommand{\msais}{MSAIS}

\newcommand{\Kepler}{{\it Kepler}}
\newcommand{\TESS}{{\it TESS}}

\providecommand{\e}[1]{\ensuremath{\, \times \, 10^{#1}}}

\newcommand{\nRVplanets}{677}
\newcommand{\nExclusions}{four}
\newcommand{\nsample}{673}

\received{October 22, 2018}
\revised{November 13, 2018}
\accepted{November 14, 2018}

\submitjournal{Publications of the Astronomical Society of the Pacific (PASP)}

\shorttitle{{\it TESS} Yield of Transits from Known Radial Velocity Exoplanets}
\shortauthors{Dalba et al.}

\begin{document}

\title{Predicted Yield of Transits of Known Radial Velocity Exoplanets from the {\it TESS} Primary and Extended Missions}

\correspondingauthor{P. A. Dalba}
\email{pdalba@ucr.edu}


\author[0000-0002-4297-5506]{Paul A. Dalba}
\affil{Department of Earth Sciences, University of California Riverside, 900 University Avenue, Riverside CA 92521, USA}

\author[0000-0002-7084-0529]{Stephen R. Kane}
\affil{Department of Earth Sciences, University of California Riverside, 900 University Avenue, Riverside CA 92521, USA}

\author[0000-0001-7139-2724]{Thomas Barclay}
\affil{NASA Goddard Space Flight Center, 8800 Greenbelt Road, Greenbelt, MD 20771, USA}
\affil{University of Maryland, Baltimore County, 1000 Hilltop Circle, Baltimore, MD 21250, USA}

\author{Jacob L. Bean}
\affil{Department of Astronomy \& Astrophysics, University of Chicago, 5640 S. Ellis Avenue, Chicago, IL 60637, USA
}

\author[0000-0002-4588-5389]{Tiago L. Campante}
\affil{Instituto de Astrof\'{i}sica e Ci\^{e}ncias do Espa\c{c}o, Universidade do Porto, Rua das Estrelas, PT4150-762 Porto, Portugal} 
\affil{Departamento de F\'{i}sica e Astronomia, Faculdade de Ci\^{e}ncias da Universidade do Porto, Rua do Campo Alegre, s/n, PT4169-007 Porto, Portugal}

\author[0000-0002-3827-8417]{Joshua Pepper}
\affil{Department of Physics, Lehigh University, 16 Memorial Drive East, Bethlehem, PA 18015, USA}

\author[0000-0003-1080-9770]{Darin Ragozzine}
\affil{Brigham Young University, Department of Physics and Astronomy, N283 ESC, Provo, UT 84602, USA}

\author{Margaret C. Turnbull}
\affil{SETI Institute, Carl Sagan Center for the Study of Life in the Universe, Off-Site: 2801
Shefford Drive, Madison, WI 53719, USA}

 
\begin{abstract}

Radial velocity (RV) surveys have detected hundreds of exoplanets through their gravitational interactions with their host stars. Some will be transiting, but most lack sufficient follow-up observations to confidently detect (or rule out) transits. We use published stellar, orbital, and planetary parameters to estimate the transit probabilities for nearly all exoplanets that have been discovered via the RV method. From these probabilities, we predict that $25.5^{+0.7}_{-0.7}$ of the known RV exoplanets should transit their host stars. This prediction is more than double the amount of RV exoplanets that are currently known to transit. The {\it Transiting Exoplanet Survey Satellite} ({\it TESS}) presents a valuable opportunity to explore the transiting nature of many of the known RV exoplanet systems. Based on the anticipated pointing of {\it TESS} during its two-year primary mission, we identify the known RV exoplanets that it will observe and predict that $11.7^{+0.3}_{-0.3}$ of them will have transits detected by {\it TESS}. However, we only expect the discovery of transits for $\sim$3 of these exoplanets to be novel (i.e., not previously known). We predict that the {\it TESS} photometry will yield dispositive null results for the transits of $\sim$125 RV exoplanets. This will represent a substantial increase in the effort to refine ephemerides of known RV exoplanets. We demonstrate that these results are robust to changes in the ecliptic longitudes of future {\it TESS} observing sectors. Finally, we consider how several potential {\it TESS} extended mission scenarios affect the number of transiting RV exoplanets we expect {\it TESS} to observe. 

\end{abstract}

\keywords{planets and satellites: detection --- surveys --- methods: statistical}


\section{Introduction} \label{sec:intro}

The vast majority of presently known exoplanets were discovered by dedicated transit surveys using space-based \citep[e.g., \Kepler,][]{Borucki2010,Twicken2016} and ground-based observatories \citep[e.g., KELT, HATNet, and WASP,][]{Pepper2003,Bakos2004,Pollacco2006}. The sheer number of stars that can be photometrically monitored in such surveys clearly overwhelms the probability of having the proper viewing geometry to witness a transit. However, the inverse proportionality between transit probability and orbital semi-major axis \citep[e.g.,][]{Beatty2008} has largely limited the exoplanet characteristics inferred from transit surveys to the inner regions of planetary systems, within a few tenths of an AU from the host star \citep[e.g.,][]{Howard2012}. 

The detection of exoplanets in radial velocity (RV) measurements of their host stars \citep[e.g.,][]{Pepe2004,Howard2010} is less biased toward short-period exoplanets than the transit method. As a result, RV surveys have revealed exoplanet characteristics at moderate orbital distances (tenths of an AU to several AU from the host star), which has greatly benefited theoretical investigations of planet formation \citep[e.g.,][]{Dawson2013,Santerne2016}.

At the intersection of the sensitivities of the transit and RV methods exists an opportunity to improve the accuracy of exoplanet demographics and properties through the validation and synthesis of independent observations. The combination of transit and RV occurrence rates better constrains the true underlying exoplanet population and identifies the biases of each method \citep[e.g.,][]{Howard2012,Wright2012,Dawson2013,Guo2017}. Furthermore, the combination of planet mass and radius---the basic properties inferred from RV and transit observations, respectively---advances the status of an exoplanet from merely a detected world, to one that can be characterized in depth. An exoplanet's bulk density (derived from planet mass and radius) provides a window to its internal composition \citep[e.g.,][]{Weiss2014}. Furthermore, mass and radius measurements are imperative for atmospheric characterization, which may place even more precise constraints on planetary interior \citep[e.g.,][]{Miller2011,Thorngren2016} and formation processes \citep[e.g.,][]{Garaud2007,Oberg2011}. 

One of the following two approaches is typically employed to obtain measurements of an exoplanet's mass and radius. Either the exoplanet is detected in a transit survey and follow-up RV observations covering the orbital phase of the planet are acquired, or the exoplanet is detected in an RV survey and follow-up photometry within the predicted transit window is acquired. Here, we consider the latter, which is potentially the riskier of the two approaches. RV exoplanet detections are biased toward edge-on orbital inclinations, but substantial photometric follow-up campaigns may nonetheless fail to measure the planet radius in the event that the exoplanet is non-transiting. In many cases, this risk factor has precluded sufficient photometric monitoring in search of transits for known RV exoplanets. This is especially true for exoplanets with relatively long orbital periods.  

Substantial efforts have previously been made to improve orbital ephemerides known exoplanets. For exoplanets discovered in transit surveys, observations of subsequent transits are typically needed to constrain transit timing variations \citep[e.g.,][]{Dalba2016,Wang2018}. For exoplanets discovered in RV surveys, multiple types of follow-up observations are usually required. Efforts to follow-up known RV exoplanets such as the Transit Ephemeris Refinement and Monitoring Survey \citep[TERMS;][]{Kane2009} acquire spectroscopic and photometric observations to search for transits and characterize host stars. TERMS and other similar efforts have successfully ruled out \citep[e.g.,][]{Kane2011c} and discovered \citep[e.g.,][]{Moutou2009} transits of a handful of RV exoplanets. However, there are simply not enough telescope resources to target all known RV exoplanets individually. Instead, this luxury can potentially be afforded by dedicated large-scale transit-hunting missions.

With the launch of the {\it Transiting Exoplanet Survey Satellite} \citep[\TESS,][]{Ricker2015} came the new opportunity to acquire light curves of known hosts of RV exoplanets. The potential yield of exoplanet discoveries from \TESS\ has been thoroughly explored \citep{Sullivan2015,Barclay2018,Ballard2018,Huang2018}, yet little effort has been focused on the yield of transits from known RV exoplanets. \citet{Yi2018} evaluated the detectability of RV exoplanets from the anticipated {\it Characterizing Exoplanets Satellite} \citep[\textit{CHEOPS};][]{Broeg2013}. However, their investigation focused on making accurate predictions of RV exoplanet radii and subsequently transit signal-to-noise ratio (SNR) to maximize the efficiency of the {\it CHEOPS} observing strategy. \citet{Yi2018} calculated single point-estimates of transit probability and provided a qualitative interpretation, but their predicted yield of {\it CHEOPS} transits was set by SNR arguments \citep{Yi2018}. An extension of the transit probability calculation to the expected yield of transits for all known RV exoplanets, including those to be observed by \TESS, remains to be done.

In this paper, we take a probabilistic approach to predicting the number of RV exoplanets that transit their host stars and the fraction of those that \TESS\ may detect. In Section \ref{sec:selection}, we explain the selection criteria used to generate our sample of known RV exoplanets. In Section \ref{sec:p_transit}, we describe the determination of all necessary exoplanet parameters and the Monte Carlo simulation of each exoplanet's transit probability\footnote{The code developed here to predict the transit probabilities of RV exoplanets will be made publicly available at \url{https://github.com/pdalba/transit_prob}.}. For each exoplanet in our sample, we report parameters describing the probability density function for its transit probability. In Section \ref{sec:expected_N}, we predict the total number of RV exoplanets that transit their host stars and offer comparison to the currently known sample of transiting RV exoplanets. In Section \ref{sec:expected_N_tess}, we refine the predicted number of transiting RV exoplanets to only those that may be observed by \TESS\ during its primary mission. We then extend this consideration of \TESS\ to several potential extended mission scenarios. In Section \ref{sec:discussion}, we discuss the influence of several simplifying assumptions that we employ in our analysis on our results. Finally, in Section \ref{sec:conclusions}, we summarize our findings.


\section{Selection of RV Exoplanets}\label{sec:selection}

All exoplanet data used here were acquired from the NASA Exoplanet Archive\footnote{Accessed 2018 August 28.} (NEA). Of the total 3778 confirmed exoplanets, we select only the \nRVplanets\ exoplanets that had ``Radial Velocity'' as the discovery method. A small fraction of this subset of exoplanets has since been found to transit their host stars. We do not exclude these exoplanets from our analysis so that we may compare the number of RV exoplanets that are currently known to transit with the number that we predict to transit based on our statistical arguments.


\section{Transit Probability}\label{sec:p_transit}

For an exoplanet on an eccentric orbit about its host star, the probability that a distant observer witnesses a transit ($p_{\rm transit}$) can be expressed as
\begin{equation} \label{eq:p-transit}
    p_{\rm transit} = \frac{(R_p + R_{\star})[1+e\cos{(\pi/2-\omega)}]}{a(1-e^2)}
\end{equation}

\noindent where $R_p$ is the planetary radius, $R_{\star}$ is the stellar radius, $a$ is the orbital semi-major axis, $e$ is the orbital eccentricity, and $\omega$ is the longitude of periastron \citep[e.g.,][]{Kane2007,Winn2010}. This relation assumes a uniform distribution for the cosine of the orbital inclination and ignores the prior probability distribution for the exoplanet mass. Therefore, when applied to RV exoplanets, Equation (\ref{eq:p-transit}) is the prior transit probability, not the posterior transit probability \citep{Ho2011,Stevens2013}. For Jupiter-like exoplanets, the prior and posterior transit probabilities are nearly equal \citep{Stevens2013}. Given that the sample of known RV exoplanets is comprised of mostly Jupiter-like exoplanets, we continue our analysis using the relation for prior transit probability (Equation \ref{eq:p-transit}). The implications of this choice are considered in Section \ref{sec:discussion}.

The NEA data have been federated from multiple sources, so some exoplanets lack one or more of the parameters needed to evaluate Equation (\ref{eq:p-transit}). Furthermore, the vast majority of exoplanets in our sample do not have reported $R_p$ values, as they either do not transit or have not been found to transit their host stars. Before evaluating transit probabilities, we give special consideration to missing parameter values and the estimation of planet radii.


\subsection{Missing Parameter Values}\label{sec:missing}

In some cases, the NEA lacks values of the physical parameters necessary to evaluate Equation (\ref{eq:p-transit}). Here, we explain our decision process regarding missing data.

If the NEA does not contain a value of the longitude of periastron, we assume $\omega = \pi/2$. If the NEA does not contain a value of orbital eccentricity, we assume $e=0$ (i.e., a circular orbit). The assumption of $e=0$ is somewhat of a simplification. The long-period nature of many of the known RV exoplanets suggests that they likely have nonzero orbital eccentricities \citep{Kipping2013a}. The assumption of $e=0$ therefore yields a lower limit for the transit probability. The resulting effect on the full set of transit probability calculations is likely minimal, but we encourage any future extensions of this work to consider physically-motivated eccentricities, such as Beta \citep{Kipping2013a} or Rayleigh \citep[e.g.,][]{Fabrycky2014} distributions.

If the NEA does not contain a value of stellar radius, we estimate its value through the following methods in the following order. First, if values of stellar surface gravity (log $g$) and stellar mass ($M_{\star}$) are available, we estimate $R_{\star}$ from the relation \citep{Smalley2005}
\begin{equation}\label{eq:smalley2005}
    \log g - \log g_{\sun} =  \log \left (  \frac{M_{\star}}{M_{\sun}} \right ) -2 \log \left (  \frac{R_{\star}}{R_{\sun}} \right ) 
\end{equation}

\noindent where the solar surface gravity is log $g_{\sun} = 4.4374$, the solar mass is $M_{\sun} = $ 1.989\e{30} kg, and the solar radius is $R_{\sun} = $ 6.96\e{8} m. Second, if values of stellar $B$-band and $V$-band magnitudes are available, we estimate $R_{\star}$ with a $B-V$ color look-up table under the assumption that the exoplanet host is a dwarf star \citep[][page 388]{Cox2000}. Third, if $B$- or $V$-band magnitude is unavailable or if the $B-V$ color is outside of the range of the look-up table, we exclude this source from further analysis (\nExclusions\ exclusions in total: HD 37124 b, HD 37124 c, HD 37124 d, and HD 41004 B b).

If the NEA does not contain a value of orbital semi-major axis, we use the stellar mass, orbital period ($P$), and planet mass ($M_p$) to estimate $a$ from Kepler's Third Law under the assumption that the orbit has an edge-on inclination. This assumption slightly biases the value of $M_p$ used in Kepler's Third Law. However, since $M_p \ll M_{\star}$ in all cases, the effect of this bias on the sum of stellar and planetary mass is negligible. Only 27 exoplanets in our sample lack an $a$ value in the NEA, and each of those contain values for $P$, $M_{\star}$, and $M_p$. Hence, none are excluded from further analysis at this step. After all exclusions, our final sample contains \nsample\ exoplanets (of a total \nRVplanets\ exoplanets).


\subsection{Estimation of Exoplanet Radii}

Although a few of the exoplanets in our sample have measured radii, we ignore this information to maintain consistency with the vast majority of RV exoplanets. We estimate $R_p$ for all of the exoplanets in our sample using the \texttt{forecaster} tool \citep{Chen2017}. \texttt{forecaster} applies probabilistic mass-radius relations to an input exoplanet mass and returns an exoplanet radius along with lower and upper uncertainties based on a posterior probability distribution \citep{Chen2017}.


\subsection{Monte Carlo Simulations of Transit Probability}\label{sec:mc}

Using measurements or estimates of each of the necessary stellar, planetary, and orbital parameters, we conduct a Monte Carlo simulation of the transit probabilities from Equation (\ref{eq:p-transit}). We begin by approximating a probability density function (PDF) for each physical parameter from which parameter values can be drawn. For parameters with symmetric errors (i.e., equal upper and lower errors), we use a normal PDF peaked at the parameter value and having a standard deviation equal to the parameter uncertainty. 

For parameters with asymmetric errors, we employ a skew-normal PDF \citep{Azzalini1985}. A skew-normal PDF is defined by three functional parameters, which control location, shape, and skewness. We obtain these functional parameters through the non-linear least-squares regression technique described by \citet[][Appendix C]{Espinoza2015}. Briefly, the physical parameter value and its upper and lower uncertainties are assumed to have been derived from the 16th, 50th, and 84th percentiles of their PDF. The residuals between these percentiles and those of a skew-normal distribution defined by the location, shape, and skewness parameters are then minimized. 

For parameters with no uncertainties provided by the NEA, we assume symmetric errors of 10\% the parameter value. This decision primarily influences stellar radii determined through Equation (\ref{eq:smalley2005}). For our sample of exoplanets, many of those lacking values of $R_{\star}$ also lack uncertainties on $M_{\star}$, despite having a listed value for $M_{\star}$. Propagating arbitrarily large errors on $M_{\star}$ through Equation (\ref{eq:smalley2005}) leads to even larger errors on $R_{\star}$, which artificially inflate the transit probability. We find that a value of 10\% error does not inflate $p_{\rm transit}$. 

\begin{figure}
    \centering
    \includegraphics[width=0.6\columnwidth]{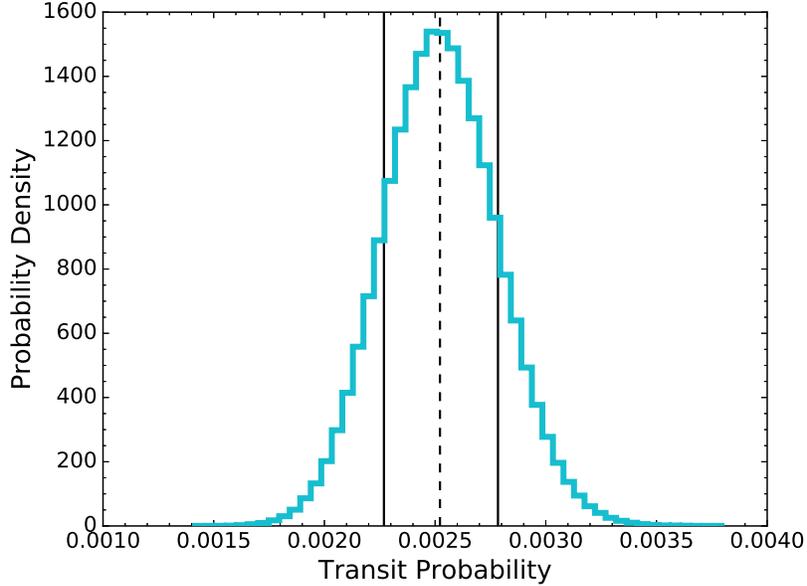}
    \caption{Normalized histogram of the transit probability ($p_{\rm transit}$), for the RV exoplanet HD 29012 b as an example, found by a Monte Carlo simulation of Equation (\ref{eq:p-transit}). The 50th percentile is marked by a dashed black line, and the 16th and 84th percentiles are marked by solid black lines. The distribution of $p_{\rm transit}$ values is slightly skewed toward higher values of $p_{\rm transit}$.}
    \label{fig:p_transit_HD29021b}
\end{figure}

We evaluate the transit probability (Equation \ref{eq:p-transit}) 1\e{6} times, by drawing parameter values from the aforementioned PDFs. Only probabilities in the range $0 \le p_{\rm transit} \le 1$ are allowed. The result is a PDF for $p_{\rm transit}$, an example of which is shown in Figure \ref{fig:p_transit_HD29021b}. The PDFs for $p_{\rm transit}$ are not necessarily normal, so we describe them using percentiles instead of the mean and standard deviation. In Table \ref{tab:all_planets}, we describe the location and scale of the $p_{\rm transit}$ PDFs for all exoplanets in our sample using the 16th, 50th, and 84th percentiles of the distributions. 

\section{Predicted Number of RV Exoplanets that Transit Their Host Stars}\label{sec:expected_N}

We use the transit probability PDFs for each exoplanet in our sample to predict the total number ($N$) of RV exoplanets that have the proper orientation to transit their host stars. First, we let $n_i$ be a random variable representing the number of transits the $i$th exoplanet contributes to the total. The value of $n_i$ can either be zero or unity. The expectation of $n_i$ is simply $\mathbb{E}[n_i] = p_{{\rm transit},i} \; n_i = p_{{\rm transit},i}$. It follows that the total number of transiting planets in the entire sample of \nsample\ RV exoplanets is described by
\begin{equation}\label{eq:N}
    N = \sum_{i=1}^{\nsample} p_{{\rm transit},i} 
\end{equation}

\begin{figure}
    \centering
    \includegraphics[width=0.6\columnwidth]{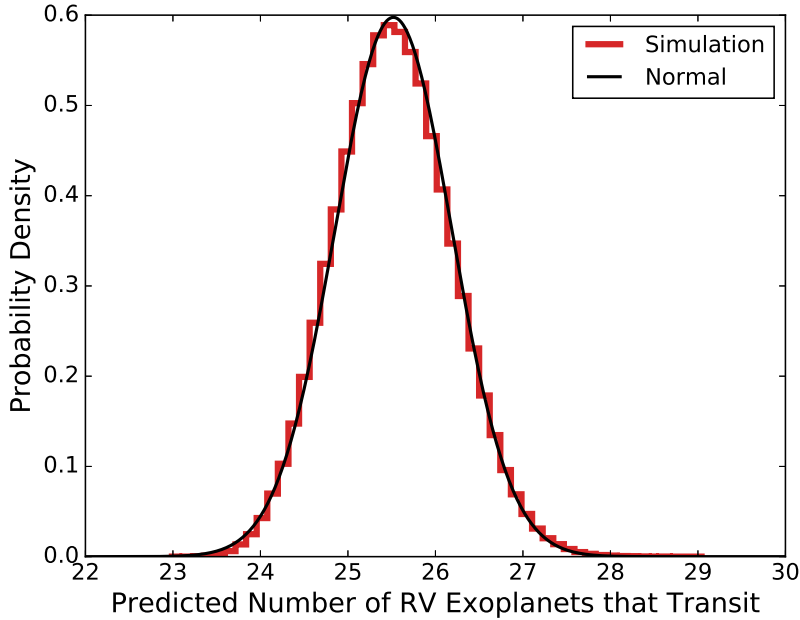}
    \caption{Normalized histogram of $N$, the predicted number of RV exoplanets in our sample that transit their host stars. The black curve is a normal distribution with the same mean and standard deviation of the distribution of $N$ values. Based on our Monte Carlo simulation of Equation (\ref{eq:p-transit}), we predict that $25.5^{+0.7}_{-0.7}$ RV exoplanets have the proper orientation to transit their host stars. 12 exoplanets within our sample are already known to transit, suggesting that there $\sim$13 more transiting RV exoplanets that have yet to be identified.}
    \label{fig:N_distribution}
\end{figure}

For each of the 1\e{6} evaluations of the transit probability, Equation (\ref{eq:N}) is also evaluated, resulting in a PDF for $N$ (Figure \ref{fig:N_distribution}). The distribution of $N$ has a mean of 25.5 exoplanets and a standard deviation of 0.66 exoplanet. The 16th, 50th and 84th percentiles of the distribution of $N$ values are 24.8, 25.5, and 26.2 exoplanets, respectively, which (to significant figures) yields a final estimate of $N=25.5^{+0.7}_{-0.7}$ exoplanets. 

How does the predicted number of RV exoplanets that transit compare to the number of RV exoplanets that have actually been found to do so? Of our sample of \nsample\ RV exoplanets, only 12 have been found to transit their host stars\footnote{The RV exoplanets already known to transit are those in the NEA with values of ``1'' in the ``{\it pl\_tranflag}'' column including HD 189733 b, HD 209458 b, HD 17156 b, 55 Cnc e, GJ 436 b, HD 97658 b, HD 149026 b, HD 219134 b, HD 219134 c, 30 Ari B b, HD 80606 b, and GJ 3470 b.}. Hence, the current number of transiting exoplanets from our sample is more than 19 standard deviations below the expectation. This suggests that additional photometric monitoring of RV exoplanets---and especially those with high transit probabilities (see Table \ref{tab:all_planets})---is likely to reveal transits. 

We consider whether the 12 known transiting RV exoplanets have relatively high transit probabilities compared to the rest of the sample. The 12 RV exoplanets currently known to transit have a mean transit probability of 11.6\%. To compare this value to the rest of the sample, we randomly draw 1\e{4} sets of 12 RV exoplanets (not including those known to transit) and record their mean transit probability. The final distribution of mean transit probabilities has a mean of 3.6\% and a standard deviation ($\sigma$) of 1.5\%. The 11.6\% mean transit probability of the known transiting RV exoplanets is more than 5$\sigma$ higher than the mean of the distribution. This suggests that the sample of RV exoplanets currently known to transit has significantly higher transit probabilities than the rest of the RV exoplanet sample. This result is not surprising as telescope resources that provide follow-up for RV exoplanets are limited and the exoplanets that are most likely to transit have been prioritized. 

Efforts to detect or rule out transits of RV exoplanets include the Transit Ephemeris Refinement and Monitoring Survey \citep[TERMS;][]{Kane2009}, which has been refining transit ephemerides and conducting photometric transit searches for RV exoplanets for nearly a decade. TERMS, and other similar efforts, have thoroughly ruled out transits for the exoplanet hosts GJ 581 \citep[e.g.,][]{LopezMorales2006,Dragomir2012b}, GJ 876 \citep[e.g.,][]{Shankland2006}, HD 114762 \citep{Kane2011c}, HD 63454 \citep{Kane2011d}, HD 192263 \citep{Dragomir2012a}, HD 38529 \citep[b planet,][]{Henry2013}, HD 130322 \citep[b planet,][]{Hinkel2015a}, 70 Vir \citep{Kane2015}, HD 6434 \citep{Hinkel2015b}, and HD 20782 \citep{Kane2016}. Less confident null detections of transits have also been reported for HD 168443 b \citep{Pilyavsky2011}, HD 37605 b \citep{Wang2012}, HD 156846 b \citep{Kane2011a}, and Proxima Cen b \citep[e.g.,][]{Kipping2017,Blank2018}. However, resources for individual efforts such as those listed here are limited. Depending on observing strategy, large-scale photometric monitoring programs may be especially helpful for detecting or ruling out transits from known RV exoplanets. The ground-based transit surveys listed above (KELT, SuperWASP, and HATNet/HATSouth) have years of photometric observations of much of the sky. However, the typical magnitudes of known RV planet hosts tend to be brighter than even the bright limit of these surveys. Space-based transit surveys, as discussed in the following section, may then provide the best opportunity to follow-up on known RV exoplanets.


\section{Predicted Number of Transiting RV Exoplanets to be Observed by \TESS}\label{sec:expected_N_tess}

\TESS\ was launched in April 2018, and is actively searching bright stars for transiting exoplanets. Over the course of its two-year primary mission, \TESS\ will search for transits across most of the sky, including in many of the currently known RV exoplanet systems. \TESS\ has four cameras positioned in a 1x4 array that provide a combined $\sim$2300 deg$^2$ field of view. \TESS\ scans approximately half of the sky in 13 sectors, each of which receives $\sim$27 days of continuous observation (at either 2- or 30-minute cadence). Depending on the ecliptic longitude and latitude of a known RV exoplanet host, it may be observed in as many as 13 consecutive sectors. 

We determine the sectors and cameras in which all \nsample\ exoplanets in our sample will be (or were) observed. We uploaded the NEA right ascension and declination for each exoplanet in our sample to the Web \TESS\ Viewing Tool\footnote{\url{https://heasarc.gsfc.nasa.gov/cgi-bin/tess/webtess/wtv.py}} \citep[WTV;][]{Mukai2017}, which estimates the pointing of {\it TESS}'s cameras using predicted spacecraft ephemerides. At the time of writing, the WTV tool was only implemented for Cycle 1 of the \TESS\ primary mission. We estimate the \TESS\ pointings for Cycle 2 by assuming that the ecliptic longitudes of the sector boundaries are the same as Cycle 1, and mirroring the ecliptic latitudes from the southern to the northern ecliptic hemispheres. The output sector(s) and camera(s) (if any) for each RV exoplanet in our sample are provided in Table \ref{tab:all_planets}.

\begin{figure*}
    \centering
    \includegraphics[width=0.85\textwidth]{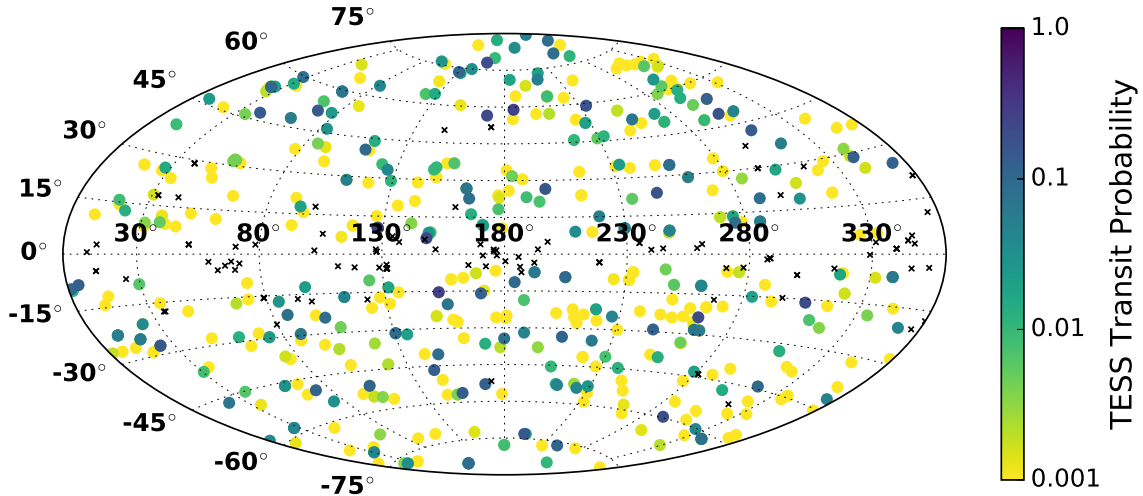}
    \caption{All known RV exoplanets in our sample displayed in ecliptic coordinates and colored by the probability that \TESS\ observes a transit ($p_{\rm transit, TESS}$). Exoplanets symbolized by black ``x'' markers are not expected to be observed by \TESS. For stars hosting multiple exoplanets, the color represents the highest $p_{\rm transit, TESS}$ value.}
    \label{fig:TESS_planets}
\end{figure*}

We explore how this assumption---and the potential alteration of {\it TESS}'s predicted pointing---affects our results by shifting the ecliptic longitudes of the Cycle 2 sector boundaries in increments of $+5^{\circ}$, up to a maximum of $+30^{\circ}$, and repeating our analysis. This amount of shifting is sufficient to give \TESS\ its original total field of view. We find that specific longitude shifts cause specific stars to receive different amounts of observation or to fall outside of {\it TESS}'s field of view (e.g., in between adjacent sectors at low ecliptic latitudes). However, variation of the ecliptic longitudes of the Cycle 2 sector boundaries do not alter the results for all RV exoplanets discussed below in a statistically significant manner. This outcome is expected because the distribution of RV exoplanets and their transit probabilities on the sky is essentially random (Figure \ref{fig:TESS_planets}). Nonetheless, this outcome is significant because it suggests that the search for transits of known RV exoplanets is resilient to changes in the ecliptic longitudes of {\it TESS}'s sector boundaries.   

Stars at different ecliptic latitudes receive different amounts of \TESS\ observation. If the $i$th exoplanet host is observed in a total of $s$ \TESS\ sectors, then the total duration ($\tau_i$) of \TESS\ observation for that exoplanet host can be approximated by $\tau_i = 27s$ days. If this exoplanet's orbital period ($P_i$) is greater than $\tau_i$, then its transit probability as observed by \TESS\ is reduced by a factor of ($\tau_i/P_i$). If instead $P_i<\tau_i$, then the limited \TESS\ baseline does not affect the probability of detecting a transit. This method of estimating transiting probabilities neglects the orbital ephemerides---except for orbital period---of our known RV exoplanets. A justification and discussion of this choice is given in Section \ref{sec:discussion}.

We repeat the Monte Carlo simulation of Section \ref{sec:mc}, but we multiply each value of $p_{{\rm transit},i}$ by $(\tau_i/P_i)$ to account for {\it TESS}'s limited observational baseline. The probability ($p_{{\rm transit, TESS},i}$) that \TESS\ observes a transit of the $i$th exoplanet in our sample becomes 
\begin{equation}\label{eq:p_transit_tess}
    p_{{\rm transit, TESS},i} = p_{{\rm transit},i} \left ( \frac{\tau_i}{P_i} \right )    
\end{equation}

\noindent where $(\tau_i / P_i)$ is forced to unity for any exoplanet with an orbital period shorter than the duration of \TESS\ observation. Values of $p_{\rm transit, TESS}$ for each exoplanet in our sample are provided in Table \ref{tab:all_planets} and are displayed in Figure \ref{fig:TESS_planets}. 

\begin{figure}
    \centering
    \includegraphics[width=0.6\columnwidth]{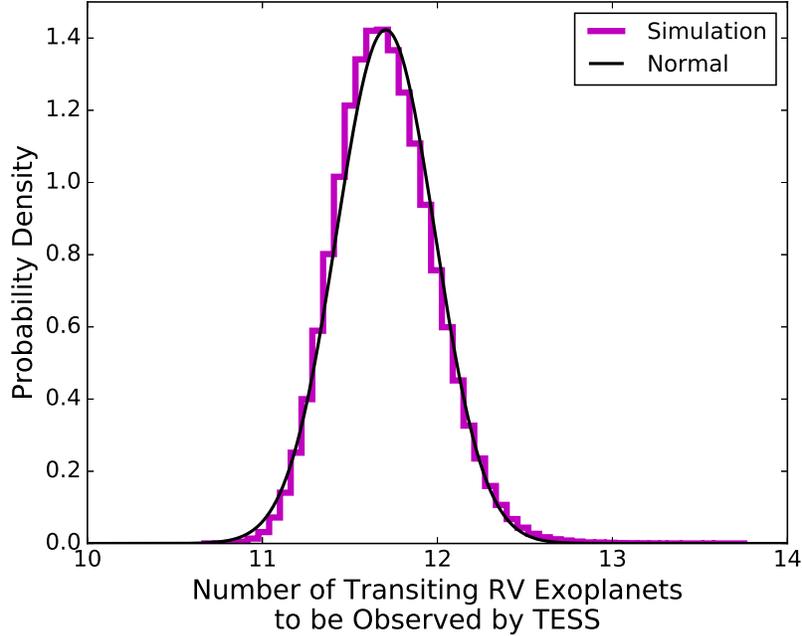}
    \caption{Normalized histogram of $N_{\rm TESS}$, the predicted number of known RV exoplanets that will observed to transit their host stars by \TESS. The solid line is a normal distribution with the same mean and standard deviation as the distribution of $N_{\rm TESS}$ values. Based on our Monte Carlo simulation of Equation (\ref{eq:p_transit_tess}), we expect that \TESS\ will observe transits from $11.7^{+0.3}_{-0.3}$ known RV exoplanets during its primary mission. Accounting for the RV exoplanets that \TESS\ will observe that are already known to transit reduces $N_{\rm TESS}$ to $\sim$3 exoplanets (see the text).}
    \label{fig:N_tess_distribution}
\end{figure}

Following Section \ref{sec:mc}, we combined the individual $p_{\rm transit, TESS}$ distributions to predict the total number ($N_{\rm TESS}$) of known RV exoplanets that will be observed to transit their host stars by \TESS\ (Figure \ref{fig:N_tess_distribution}). We find that $N_{\rm TESS} = 11.7^{+0.3}_{-0.3}$ exoplanets, meaning that we predict that \TESS\ will observe transits of 11 or 12 known RV exoplanets in our sample during its primary mission. 

We note that our full sample of \nsample\ RV exoplanets includes 12 that have previously been found to transit their host stars. Of these 12 exoplanets, 8 (HD 189733 b, HD 209458 b, HD 17156 b, 55 Cnc e, GJ 436 b, HD 149026 b, HD 219134 b, and HD 219134 c) are expected to have \TESS\ observational baselines that are equal to or longer than their orbital periods. Hence, these 8 exoplanets are essentially guaranteed to be observed in transit by \TESS. Two of the remaining four RV exoplanets already known to transit (GJ 3470 b and HD 97658 b) are not expected to be observed by \TESS. GJ 3470 b has an ecliptic latitude of $\sim-5^{\circ}$, placing it outside of {\it TESS}'s anticipated field of view. HD 97658 b has an ecliptic latitude of $\sim+19^{\circ}$, but it falls in a field of view gap between adjacent sectors\footnote{If the actual Cycle 2 sector boundaries are defined differently than we have assumed here (in Section \ref{sec:expected_N_tess})---such that HD 97658 is in the field of view of \TESS---the transit of HD 97658 b (with $P \approx 9.5$ days) will be observed.}. The other two RV exoplanets already known to transit (30 Ari B b and HD 80606 b) have ($\tau/P$) ratios of 0.08 and 0.24, respectively, so their respective \TESS\ transit probabilities $p_{\rm transit, TESS}$ also equal 0.08 and 0.24.

The contribution of RV exoplanets that are already known to transit can be removed from the predicted value of $N_{\rm TESS}$ by subtracting 8 to account for the ``guaranteed'' transits and by subtracting an additional 0.32 to account for the $p_{\rm transit, TESS}$ values of 30 Ari B b and HD 80606 b. In summary, during its primary mission, we predict that \TESS\ will discover transits of $\sim3$ known RV exoplanets that were not previously known to transit. 

The low value of $N_{\rm TESS}$ can be attributed to the long-period nature of the sample of known RV exoplanets, especially when contrasted with the relatively short observational baseline that most of the \TESS\ fields will receive. The median and mean orbital periods of our exoplanet sample are 381 days and 867 days, respectively, demonstrating the distribution is significantly skewed toward long periods. Of the \nsample\ known RV exoplanets in our sample, 125 have $(\tau/P) \ge 1$, meaning that the \TESS\ baseline does not reduce the probability of observing a transit. For these exoplanets, \TESS\ will be able to confidently refute or confirm the transiting nature of the system. This will represent a substantial contribution to ephemeris refinement efforts such as TERMS \citep{Kane2009}, and one that would otherwise necessitate a tremendous amount of ground-based resources. {\it TESS}'s contribution to ephemeris refinement will be especially valuable for exoplanets with long periods. The long-period end of 125 exoplanets with $(\tau/P) \ge 1$ includes HD 40307 g ($P\approx 198$ days, $p_{\rm transit, TESS} \approx 0.95$\%), HD 65216 c ($P\approx153$ days, $p_{\rm transit, TESS} \approx 0.83$\%), and HD 156279 b ($P\approx 131$ days, $p_{\rm transit, TESS} \approx 3.5$\%) among others (see Table \ref{tab:all_planets}).


\subsection{Signal-to-noise Considerations for \TESS}

Our prediction for the number of known RV exoplanets for which \TESS\ will discover transits has so far lacked any consideration of the depths of the potential transits, the brightness of the host stars, and the anticipated precision of the \TESS\ photometry. Here, we briefly consider these factors in relation to detecting transits of RV exoplanets in \TESS\ observations.

\begin{figure}
    \centering
    \includegraphics[width=0.6\columnwidth]{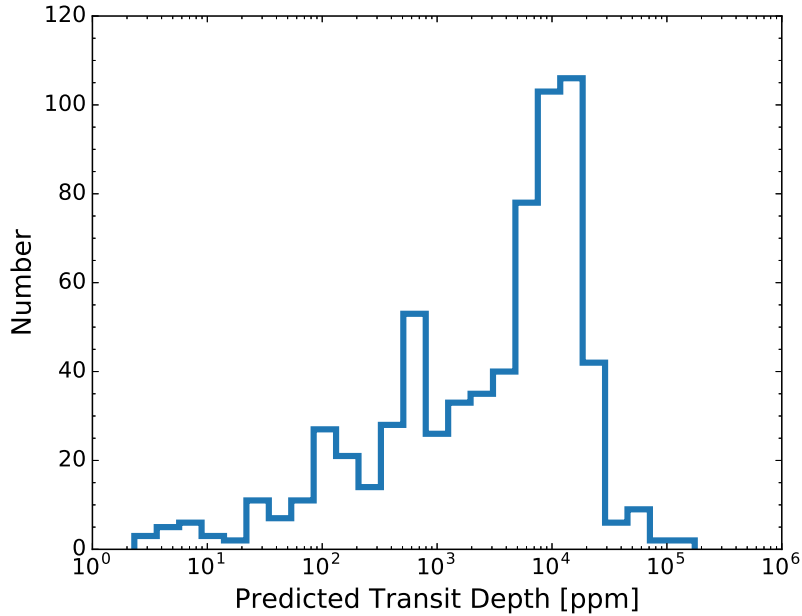}
    \caption{Histogram of the predicted transit depth as approximated by $(R_p/R_{\star})^2$ for the known RV exoplanets in our sample. $R_p$ values are estimated using measured exoplanet masses and the mass-radius relations of \citet{Chen2017}. $R_{\star}$ values are either measured or estimated as described in Section \ref{sec:missing}. The majority of this sample have predicted transit depths of $\sim$1\%, suggesting that the sample is composed mainly of giant exoplanets.}
    \label{fig:transit_depth}
\end{figure}

In Figure \ref{fig:transit_depth}, we show the predicted distribution of transit depths for all RV exoplanets in our sample. We approximate transit depth as $(R_p/R_{\star})^2$, where $R_p$ and $R_{\star}$ are determined as described in Section \ref{sec:p_transit}. The distribution of the predicted transit depths spans $\sim$2--2\e{5} parts per million (ppm) and peaks near 1\e{4} ppm, which is the equivalent of Jupiter transiting the Sun. This is expected as our sample of RV exoplanets is known to contain mostly giant planets. 

It is expected that 1\% precision should be readily achievable for most \TESS\ targets. However, we investigate the \TESS\ detection efficiency of the known RV exoplanet sample more carefully by estimating the signal-to-noise ratio (SNR$_{\rm TESS}$) of their potential transit signals. First, we assume circular, edge-on transits so that the transit duration ($t_{\rm dur}$) is given by \citep{Winn2010}
\begin{equation}\label{eq:t_dur}
    t_{\rm dur} = \frac{P}{\pi} \arcsin{\left [ \frac{R_{\star}}{a}  \left ( 1 + \frac{R_p}{R_{\star}}\right )      \right ] } \;.
\end{equation}

\noindent The transit duration is evaluated with the same parameters used previously to estimate transit probability. Then, for a transit signal containing $\mathcal{N}$ transits, we estimate SNR$_{\rm TESS}$ following \citet{Barclay2018}:
\begin{equation}\label{eq:snr}
    {\rm SNR}_{\rm TESS} = \frac{(R_p/R_{\star})^2}{\sigma_{\rm TESS}}  \frac{\sqrt{\mathcal{N} t_{\rm dur}}}{(1+d)}
\end{equation}

\noindent where $d$ is the dilution of the transit signal due to nearby sources and $\sigma_{\rm TESS}$ is the photometric noise model for \TESS\ from Figure 14 of \citet{Sullivan2015}. Predicted values of $d$ are acquired from the \TESS\ Candidate Target List \citep{Stassun2018} in the ``\texttt{contratio}'' column. In Equation (\ref{eq:snr}), $t_{\rm dur}$ has units of hr and $\sigma_{\rm TESS}$ has units of hr$^{1/2}$. We estimate the photometric noise for our sources using their \TESS-band magnitudes as a proxy for the standard Johnson--Cousins $I_C$ band. 

In calculating SNR$_{\rm TESS}$, we consider the total amount of observing time that a star is expected to receive by defining $\mathcal{N}$ as the integer value of $\tau/P$ (rounding down). For $0<(\tau/P)<1$, we ascribe $\mathcal{N}=1$, as if a single transit is detected. The purpose of this calculation is not to consider the probability of such a transit detection, but instead to compare its signal to the noise assuming a transit was to occur. 

\begin{figure}
    \centering
    \includegraphics[width=0.6\columnwidth]{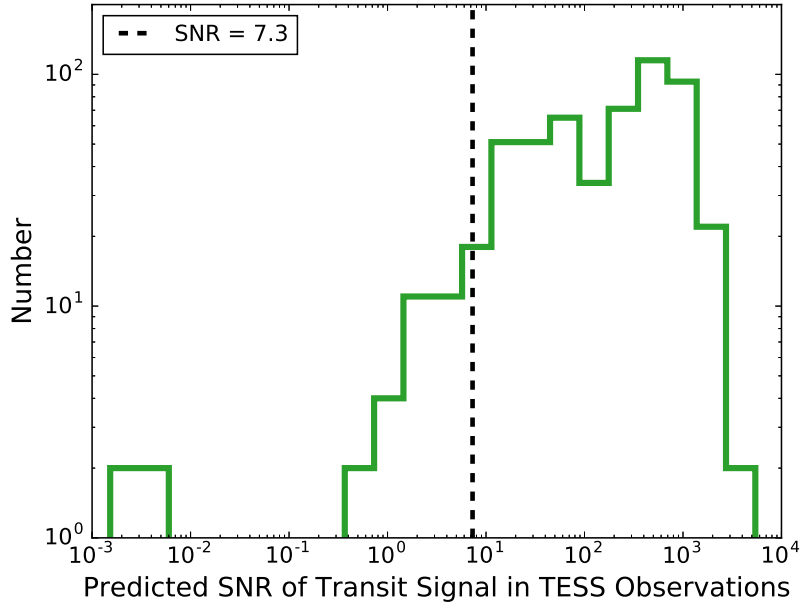}
    \caption{Histogram of the predicted SNR of the potential transit signals of the known RV exoplanets. Only exoplanet hosts that are expected to be observed by \TESS\ (i.e., $\tau>$ 0 days) are included in this distribution. The vertical dashed line is plotted at SNR=7.3, the detection threshold adopted by \citet{Sullivan2015}. 93\% of the plotted distribution has SNR values above 7.3, suggesting that photometric precision is unlikely to reduce {\it TESS}'s ability to recover transits of known RV exoplanets.}
    \label{fig:transit_snr}
\end{figure}

Figure \ref{fig:transit_snr} shows the predicted distribution of SNR$_{\rm TESS}$ for the all of the known RV exoplanets that are expected to be observed by \TESS. We compare this distribution to the nominal detection threshold of 7.3, which was derived by \citet{Sullivan2015} for \TESS\ following a similar calculation for the original \Kepler\ mission \citep{Jenkins2010}. Of the known RV exoplanets expected to be observed by \TESS, 93\% are predicted to have SNR$_{\rm TESS} \ge 7.3$. Alternatively, eight exoplanets are expected to have SNR$_{\rm TESS} < 1$. These include the five known exoplanets orbiting GJ 667 C (which suffers from extreme flux dilution) and three others orbiting host stars with anomalously large radii values listed in the NEA (i.e., HD 208527, BD+20 2457, and HD 96127). Since the vast majority of RV exoplanets would likely produce transit signals that are significantly larger than {\it TESS}'s detection threshold, it is unlikely that the yield of transits from the RV exoplanets would be significantly reduced due to insufficient photometric precision.

\begin{deluxetable*}{lccccccc}
    \tablecaption{Transit probabilities and predicted \TESS\ observation details for the known RV exoplanets in our sample} 
    \tablecolumns{8}
    \tablewidth{0pt}
    \tablehead{ 
         &  & & & \multicolumn{2}{c}{\TESS\ Cycle 1} & \multicolumn{2}{c}{\TESS\ Cycle 2\tablenotemark{c}}\\
        \cline{5-8}
        \colhead{Exoplanet Name} & 
        \colhead{TIC ID\tablenotemark{a}} &
        \colhead{$p_{\rm transit}$\tablenotemark{b}} &
        \colhead{$p_{\rm transit, TESS}$\tablenotemark{b}} &
        \colhead{Sector(s)} & 
        \colhead{Camera(s)} &
        \colhead{Sector(s)} & 
        \colhead{Camera(s)} 
        }
    \startdata 
        NGC 2682 Sand 364 b & 294684871 & 0.5$^{+0.5}_{-0.5}$ & \nodata & \nodata & \nodata & \nodata & \nodata\\
        BD+20 2457 b & 99734092 & 0.4$^{+0.4}_{-0.4}$ & 0.03$^{+0.03}_{-0.03}$ & \nodata & \nodata & 8 & 1\\
        55 Cnc e & 332064670 & 0.291$^{+0.005}_{-0.004}$ & 0.291$^{+0.005}_{-0.004}$ & \nodata & \nodata & 7 & 1\\
        HD 240237 b & 323919676 & 0.29$^{+0.1}_{-0.08}$ & 0.021$^{+0.007}_{-0.006}$ & \nodata & \nodata & 3,4 & 3,3\\
        HD 102956 b & 428673146 & 0.28$^{+0.03}_{-0.03}$ & \nodata & \nodata & \nodata & \nodata & \nodata\\
        HD 86081 b & 27073220 & 0.27$^{+0.03}_{-0.03}$ & 0.27$^{+0.03}_{-0.03}$ & 8 & 1 & \nodata & \nodata\\
        24 Boo b & 441712711 & 0.26$^{+0.03}_{-0.02}$ & 0.26$^{+0.03}_{-0.02}$ & \nodata & \nodata & 9,10 & 3,3\\
        HD 96127 b & 268634769 & 0.2$^{+0.1}_{-0.1}$ & 0.01$^{+0.005}_{-0.005}$ & \nodata & \nodata & 8 & 2\\
        HD 88133 b & 54237857 & 0.23$^{+0.03}_{-0.03}$ & 0.23$^{+0.03}_{-0.03}$ & \nodata & \nodata & 8 & 1\\
        HD 24064 b & 286923464 & 0.22$^{+0.04}_{-0.03}$ & 0.022$^{+0.004}_{-0.003}$ & \nodata & \nodata & 5,6 & 2,2\\
        HD 118203 b & 120960812 & 0.21$^{+0.02}_{-0.02}$ & 0.21$^{+0.02}_{-0.02}$ & \nodata & \nodata & 9 & 3\\
        GJ 674 b & 20096620 & 0.2$^{+0.2}_{-0.2}$ & 0.2$^{+0.2}_{-0.2}$ & 12 & 1 & \nodata & \nodata\\
        HD 1690 b & 37763102 & 0.21$^{+0.07}_{-0.06}$ & 0.01$^{+0.003}_{-0.003}$ & 3 & 1 & \nodata & \nodata\\
        WASP-94 B b & 68418284 & 0.21$^{+0.02}_{-0.02}$ & 0.21$^{+0.02}_{-0.02}$ & 1 & 1 & \nodata & \nodata\\
        HD 212301 b & 161314759 & 0.2$^{+0.02}_{-0.02}$ & 0.2$^{+0.02}_{-0.02}$ & 1,13 & 3,3 & \nodata & \nodata\\
        HD 13189 b & 63564248 & 0.2$^{+0.03}_{-0.03}$ & \nodata & \nodata & \nodata & \nodata & \nodata\\
        BD+20 2457 c & 148563075 & 0.2$^{+0.2}_{-0.2}$ & 0.008$^{+0.009}_{-0.008}$ & \nodata & \nodata & 8 & 1\\
        \multicolumn{1}{c}{\vdots} & \vdots & \vdots & \vdots & \vdots & \vdots & \vdots & \vdots 
    \enddata
    \tablecomments{Table \ref{tab:all_planets} will be published in its entirety in the machine-readable format as Supplementary Data.}
    \tablenotetext{a}{Unique identifier from the \TESS\ Input Catalog \citep[TIC,][]{Stassun2018}.}
    \tablenotetext{b}{Transit probabilities and uncertainties are approximated using the 16th, 50th, and 84th percentiles of their probability distributions.}
    \tablenotetext{c}{Cycle 2 sectors and cameras are determined using the ecliptic longitudes and latitudes of Cycle 1 mirrored to the northern ecliptic hemisphere (see the text).}
    \label{tab:all_planets}
\end{deluxetable*}


\subsection{Potential \TESS\ Extended Missions}

The long-term stability of {\it TESS}'s lunar resonance orbit \citep{Gangestad2013} coupled with the anticipated use of spacecraft fuel makes an extended mission a realistic possibility. Here, we consider how the predicted number of transits of known RV exoplanets increases for three potential extended mission scenarios. The scenarios discussed here are by no means an exhaustive list of all possibilities. In this discussion, we continue to only use orbital periods and ignore the other orbital ephemerides when estimating \TESS\ transit probabilities (see Section \ref{sec:discussion}). 

The first strategy for an extended mission involves obtaining a long observational baseline over a relative small portion of the sky. Scenarios such as ``\texttt{pole}'' \citep{Bouma2017} or ``\texttt{C3PO}'' \citep{Huang2018} achieve this by pointing \TESS\ toward either of the ecliptic poles for timescales of a year or longer. As displayed in Figure \ref{fig:TESS_planets}, the known RV exoplanets are distributed approximately evenly across the sky. Therefore, focusing observations on a small area would likely minimize the number of transits detected from known RV exoplanets. This strategy perhaps allows for the recovery of transits from known RV exoplanets with long orbital periods, up to the observational baseline (see Section \ref{sec:conclusions}). However, the transit probability for long-period exoplanets is inherently lower, meaning that a transit is less likely to occur regardless of how long the host star is observed. 

On the other hand, the second strategy for an extended mission involves repeating the nominal primary mission: dividing the ecliptic hemisphere into 13 sectors, observing each sector for $\sim$27 days, and repeating for the opposite ecliptic hemisphere \citep[e.g., ``\texttt{hemi},''][]{Bouma2017,Huang2018}. For the $i$th known RV exoplanet, this strategy yields an extended observational baseline $\tau_{{\rm ext},i}$ that satisfies 
\begin{equation}
    \tau_{{\rm ext},i} = (1 + f_{{\rm ext},i})\tau_i
\end{equation}

\noindent where $f_{{\rm ext},i}$ is the number of times during the extended mission that \TESS\ surveys the ecliptic hemisphere containing the $i$th exoplanet. By substituting $\tau_{{\rm ext},i}$ into Equation (\ref{eq:p_transit_tess}), we find that $p_{\rm transit,TESS}$ increases by as much as a factor of $(1 + f_{{\rm ext},i})$, so long as $(\tau_{{\rm ext},i}/P_i) \le 1$. If this extended mission strategy is employed for only one year in the southern ecliptic hemisphere, then the total value of $N_{\rm TESS}$ (including both the primary and extended missions) would increase to $12.2^{+0.3}_{-0.3}$ exoplanets. The same calculation for a one-year extended mission facing the northern ecliptic hemisphere yields $N_{\rm TESS} = 12.6^{+0.3}_{-0.3}$ exoplanets. Following this strategy for a six-year extended mission (three years facing south, three years facing north) would yield $N_{\rm TESS} = 15.0^{+0.3}_{-0.3}$. We find that, on average, the ``\texttt{hemi}'' extended mission scenario leads to the discovery of transits for roughly one RV exoplanet per year of the extended mission. 

The third strategy for an extended mission involves observing both ecliptic hemispheres over the course of one year using the same pointing as the nominal primary mission, but reducing the sector baseline by 50\%, to $\sim$14 days each \citep[e.g.,``\texttt{allsky},''][]{Bouma2017,Huang2018}. For the $i$th known RV exoplanet, this strategy yields an extended observational baseline $\tau_{{\rm ext},i}$ that satisfies 
\begin{equation}
    \tau_{{\rm ext},i} = (1 + 0.5f_{{\rm ext},i})\tau_i  \;,
\end{equation}

\noindent which only differs from the ``\texttt{hemi}'' strategy if the extended mission lasts for an odd number of years. In the event of a single-year extended mission that follows the ``\texttt{allsky}'' strategy, we estimate that $N_{\rm TESS}$ would increase to $12.5^{+0.3}_{-0.3}$.

In the interest of detecting transits from known RV exoplanets that are currently not known to transit, \TESS\ extended mission scenarios that provide broad sky coverage, as opposed to long observational baselines, are preferred. For a single or multi-year extended mission, repeating the nominal primary mission pointing with the same sector durations or reduced sector durations produce similar results. Although the predicted number of transiting RV exoplanets rises slowly as a function of extended mission duration, the number of RV exoplanets for which we can rule out transits will increase more rapidly. These dispositive null results will comprise a valuable portion of the science return for known RV exoplanets from the \TESS\ mission.


\section{Discussion}\label{sec:discussion}

In the following sections, we describe several of the simplifying assumptions we employed in our prediction of the yield of transiting RV exoplanets. We also evaluate the significance of these assumptions and any effects they may have had on the results.


\subsection{Coplanarity of Multi-planet Systems}\label{sec:coplanarity}

Throughout this work, the transit probability of each exoplanet is estimated independently. That is, transit probabilities of exoplanets in multi-planet systems do not take into account the coplanarity of the system or the exoplanets' mutual inclinations. This simplifying assumption merits mention as our sample consists of 100 multi-planet systems comprising 254 total exoplanets. A more accurate approach would be to convolve the transit probabilities within a system of exoplanets with the likelihood of coplanarity \citep{Brakensiek2016}. 

Depending on a particular system's properties, considering coplanarity may either increase or decrease the transit probabilities of the members of that system. When considered for the full sample, we do not expect coplanarity to significantly alter the predicted number ($N$) of RV exoplanets that transit or the predicted number ($N_{\rm TESS}$) that \TESS\ is expected to observe. However, our choice to ignore coplanarity should be considered when interpreting the transit probabilities of exoplanets belonging to multi-planets systems as listed in Table \ref{tab:all_planets}.


\subsection{Orbital Inclination Bias of the RV Method}\label{sec:inclination}

As is true for the transit method, the detection efficiency of the RV method of exoplanet discovery increases for more edge-on orbital inclinations. This means that the sample of RV exoplanets discussed in this work is biased toward edge-on orbits. The calculation of transit probabilities according to Equations (\ref{eq:p-transit}) and (\ref{eq:p_transit_tess}) ignore this bias in favor of the simplicity offered by the assumption of random inclinations. Therefore, the predictions for the number ($N$) of RV exoplanets that transit and the number ($N_{\rm TESS}$) of RV exoplanets that \TESS\ should see in transit are lower limits. 

There is an additional subtlety to the inclination bias of the RV method. For a given RV precision, a more massive exoplanet is less affected by the inclination bias than a less massive exoplanet because the RV detection efficiency also scales with planet mass. Therefore, less massive (e.g., several $M_{\earth}$) RV exoplanets are more likely to have edge-on inclinations than more massive (e.g., several hundred $M_{\earth}$) exoplanets. This subtle point further demonstrates that our predictions for $N$ and $N_{\rm TESS}$ are lower limits. However, low mass exoplanets comprise only a minor portion of our sample of RV exoplanets, so this bias has a similarly minor effect.


\subsection{Prior versus Posterior Transit Probability for RV Exoplanets}\label{sec:coplanarity}

The transit probability of an exoplanet detected via the RV method depends on the prior probability density of exoplanet mass as well as prior probability density of orbital inclination \citep[e.g.,][]{Ho2011}. Because Equation (\ref{eq:p-transit}) does not consider the \textit{a priori} mass distribution of exoplanet, it does not represent the \textit{a posteriori} transit probability of an RV exoplanet. \citet{Stevens2013} provided a thorough investigation of posterior transit probabilities that included the derivation of multiplicative factors useful for determining posterior transit probability from prior transit probability. The factors, however, required the assumption of upper limits for exoplanet masses, which we do not have for our sample of known RV exoplanets. Instead of introducing uncertainty through the assumption of mass upper limits, we chose to accept the approximation of Equation (\ref{eq:p-transit}) as the RV exoplanet transit probability. 

The impact of this choice can be evaluated qualitatively using Table 2 of \citet{Stevens2013}. Dividing our sample of RV exoplanets into mass regimes based on their measured minimum masses, we find that 12\% are ``Earth/Super-Earths'' (0.1 $M_{\earth}$ -- 10 $M_{\earth}$), 13\% are ``Neptunes'' (10 $M_{\earth}$ -- 100 $M_{\earth}$), 48\% are ``Jupiters'' (100 $M_{\earth}$ -- 10$^3$ $M_{\earth}$), 23\% are ``Super-Jupiters'' (10$^3$ $M_{\earth}$ -- 13 $M_{\rm Jupiter}$), and 4\% are ``Brown Dwarfs'' (13 $M_{\rm Jupiter}$ -- 0.07 $M_{\sun}$). The posterior transit probabilities for the Earth/Super-Earth, Neptune, and Super-Jupiter regimes (i.e., 48\% of our total sample) are larger than the prior transit probabilities by 12--19\% on average. The posterior transit probabilities for the Jupiter regime (i.e., 48.0\% of our total sample) are only 1\% smaller than the prior transit probabilities on average. The posterior transit probabilities for the Brown Dwarf regime (i.e., 4\% of our total sample) are 14\% smaller than the prior transit probabilities on average. Had we included the \textit{a priori} distribution of exoplanet masses in the calculation of transit probability, the vast majority of transit probabilities would have either stayed nearly the same or increased. This caveat again demonstrates that our predictions for $N$ and $N_{\rm TESS}$ are lower limits.


\subsection{Ignoring RV Exoplanet Ephemerides}

In determining the fraction of an exoplanet's orbital phase sampled by \TESS\ observation, we neglect all ephemeris information except for orbital period. This choice is motivated by the lack of precision in the current ephemerides of most RV exoplanets. Figure 2 of \citet{Kane2009} demonstrated that, for orbital periods longer than $\sim$100 days, the size of the exoplanet's transit window was comparable to its orbital period. Hence, the positions of those exoplanets in their orbits were essentially unknown. In the intervening 10 years since that study, most of the RV exoplanets have not received follow-up observation, so the transit windows have become even longer.  Only a small handful of RV exoplanets (e.g., the TERMS targets mentioned in Section \ref{sec:expected_N}) have been observed sufficiently frequently and recently to have precise ephemerides that enable a meaningful comparison between the transit windows and the dates of the \TESS\ sectors. Therefore, for consistency, we have assumed that the portions of all exoplanet orbits sampled by \TESS\ observations are random.


\section{Summary and Conclusions}\label{sec:conclusions}

We use published stellar, orbital, and planetary parameters to estimate the transit probabilities for nearly all exoplanets that have been discovered via the RV method. Monte Carlo simulations of these transit probabilities enable the prediction that $25.5^{+0.7}_{-0.7}$ of the RV exoplanets in our sample have the proper orbital inclination to transit their host stars (Figure \ref{fig:N_distribution}). At the time of writing, the number of exoplanets in our sample that are known to transit is only 12, which is more than 19 standard deviations below the expectation. Hence, we predict that follow-up photometric observations of known RV exoplanets are likely to yield transit discoveries. 

The ability of the transit-hunting \TESS\ mission to observe transits of known RV exoplanets is also considered. The vast majority (93\%) of RV exoplanets have predicted transit signals with SNR greater than the nominal detection threshold of 7.3 \citep{Sullivan2015}, suggesting that \TESS\ is a suitable observatory for follow-up (Figure \ref{fig:transit_snr}). We identify the known RV exoplanets that \TESS\ is likely to observe (Figure \ref{fig:TESS_planets}) and recalculate their transit probabilities to account for {\it TESS}'s finite observational baseline. Monte Carlo simulations of these updated transit probabilities enable the prediction that \TESS\ will observe transits for $11.7^{+0.3}_{-0.3}$ of the known RV exoplanets in our sample (Figure \ref{fig:N_tess_distribution}). Of these, only $\sim$3 exoplanets will likely not have been known to transit previously. We find that our prediction is robust to changes in the ecliptic longitudes of {\it TESS}'s sector boundaries. 

We also expect \TESS\ to be able to confidently rule out transits for $\sim$125 known RV exoplanets that have orbital periods that are less than or equal to the anticipated baselines of \TESS\ observations. Transit probabilities and predicted \TESS\ observation information for each exoplanet in our sample are provided in Table \ref{tab:all_planets}.

We consider several possible scenarios for a \TESS\ extended mission in regard to following up known RV exoplanets. The distribution of known RV exoplanets on the sky is essentially even, so extended missions that emphasize all-sky coverage will maximize the number of orbital ephemerides constraints for known RV exoplanets. We find that a multi-year extended mission with an all-sky observing strategy can increase the total number of RV exoplanets that are found to transit, but only by roughly one exoplanet per year. This increase is not strongly dependent on the specific all-sky observing strategy (i.e., repeating the strategy of the primary mission or observing the full sky each year). 

Alternatively, an extended mission scenario that focuses on a smaller portion of sky (e.g., an ecliptic pole) in order to acquire a longer baseline would also complement efforts to follow-up known RV exoplanets. Indeed, 23 of the known RV exoplanets in our sample have ecliptic latitudes within 13$^{\circ}$ of a pole, and 13 of these have orbital periods longer than 100 days. Detecting a transit of a known long-period RV exoplanet would be an exceptionally valuable, albeit rare, discovery. Having a measured mass and radius, this exoplanet could extend relations of planet density versus stellar insolation and potentially serve as a solar system analog. During a subsequent transit, its transmission spectrum could also be observed, an endeavor that would likely prove fruitful for terrestrial \citep[e.g.,][]{Robinson2018} or giant \citep[e.g.,][]{Dalba2015} long-period exoplanets alike. The observational strategy employed by the \TESS\ primary mission is largely geared toward short-period exoplanets. This is evident by the relatively small number of known RV exoplanets that we predict \TESS\ will observe in transit. An extended mission that focuses on longer period exoplanets, both new and previously known, would likely yield a valuable and complementary science return. 


\acknowledgments

We wish to thank the anonymous referee for thoughtful suggestions that improved this work.

This research made use of the NASA Exoplanet Archive, which is operated by the California Institute of Technology, under contract with the National Aeronautics and Space Administration under the Exoplanet Exploration Program. 

Some of the data presented in this paper were obtained from the Mikulski Archive for Space Telescopes (MAST). STScI is operated by the Association of Universities for Research in Astronomy, Inc., under NASA contract NAS5-26555.

T. L. Campante acknowledges support from the European Union's Horizon 2020 research and innovation program under the Marie Sk\l{}odowska-Curie grant agreement No. 792848.

Funding for the TESS mission is provided by NASA's Science Mission directorate.

P. A. Dalba and S. R. Kane acknowledge financial support from the NASA Exoplanet Research Program through grant 17-XRP17\_2-0148.


\vspace{5mm}

\software{\texttt{forecaster} \citep{Chen2017}, \texttt{tvguide} \citep{Mukai2017}, \texttt{astropy} \citep{astropy2013}}

\end{document}